\documentclass[12pt, draftclsnofoot, onecolumn]{IEEEtran}
\usepackage{subfigure}
\PassOptionsToPackage{subfigure}{tocloft}
\usepackage{CJK}
\usepackage{algorithm}
\usepackage{algorithmic}
\usepackage{amsmath}
\usepackage{amssymb}
\usepackage{psfrag}
\usepackage{graphicx}
\usepackage{epstopdf}
\usepackage{multirow}
\usepackage{stfloats}
\usepackage{cite}
\usepackage{color}
\usepackage[normalem]{ulem}
\usepackage{arydshln}
\usepackage{enumerate}
\usepackage{bbm}
\usepackage{setspace}
\usepackage{times,amsmath,amsthm,color,amssymb,graphicx,epsfig,algorithm,algorithmic,bm,multirow,cite}

\theoremstyle{definition}

\newtheorem{proposition}{Proposition}

\newtheorem{remark}{Remark}

\newcommand{\Tr}{\mathop{\mathrm{Tr}}}
\newcommand{\calD}{\mathcal{D}}
\newcommand{\Rank}{\mathop{\mathrm{Rank}}}

\floatname{algorithm}{Algorithm}

\title{Symbiotic Radio: A New Communication Paradigm for Passive Internet-of-Things}

\author{Ruizhe Long, Huayan Guo, Gang Yang, Ying-Chang Liang, \IEEEmembership{Fellow, IEEE}, and Rui Zhang, \IEEEmembership{Fellow, IEEE}

\thanks{This paper has been presented in part at the IEEE Global Communications Conference, Singapore, in 2017 \cite{Long2017Transmit}.}

\thanks{This work was supported in part by the National Natural Science Foundation of China under Grants 61601100, 61628103, 61631005, and 61571100.}

\thanks{R. Long, H. Guo, and G. Yang are with the National Key Laboratory of Science and Technology on Communications, and with the Center for Intelligent Networking and Communications (CINC), University of Electronic Science and Technology of China (UESTC), Chengdu 611731, China (e-mails: ruizhelong@gmail.com, guohuayan@pku.edu.cn, yanggang@uestc.edu.cn).}

\thanks{Y.-C. Liang is with the Center for Intelligent Networking and Communications (CINC), University of Electronic Science and Technology of China (UESTC), Chengdu 611731, China (e-mail: liangyc@ieee.org).}

\thanks{R. Zhang is with Department of Electrical and Computer Engineering, National University of Singapore, Singapore (e-mail: elezhang@nus.edu.sg).}

}

\begin{document}

\maketitle
\begin{abstract}
In this paper, a novel technique, called symbiotic radio (SR), is proposed for passive Internet-of-Things (IoT), in which a backscatter device (BD) is integrated with a primary transmission. The primary transmitter is designed to assist the primary and BD transmissions, and the primary receiver decodes the information from the primary transmitter as well as the BD. We consider a multiple-input single-output (MISO) SR and the symbol period for BD transmission is designed to be either the same as or much longer than that of the primary system, resulting in parasitic or commensal relationship between the primary and BD transmissions. We first derive the achievable rates for the primary system and the BD transmission.
Then, we formulate two transmit beamforming optimization problems, i.e., the weighted sum-rate maximization problem and the transmit power minimization problem, and solve these non-convex problems by applying semi-definite relaxation technique. In addition, a novel transmit beamforming structure is proposed to reduce the computational complexity of the solutions. Simulation results show that when the BD transmission rate is properly designed, the proposed SR not only enables the opportunistic transmission for the BD via energy-efficient passive backscattering, but also enhances the achievable rate of the primary system by properly exploiting the additional signal path from the BD.
\end{abstract}

\section{Introduction}
\label{sec:intro}


Internet-of-things (IoT), which aims at realizing ubiquitous connectivity for massive devices in a wireless manner \cite{Atzori2010Internet,stankovic2014research}, is one of the major applications of the forthcoming fifth generation (5G) and future wireless networks. Due to the exponential growth of the number of IoT devices, substantial amount of energy and radio spectrum resources is required to support such massive connections.

The lack of sufficient radio spectrum is one of the bottlenecks to the success of IoT. According to \cite{UniEurope2016Identification}, around 76 GHz spectrum resources are needed to accommodate the massive IoT connections if exclusive spectrum is allocated. While cognitive radio (CR) technology can be used to support the shared spectrum access for IoT \cite{Liang2011Cognitive,Khan2017Cognitive,Zhang2018Spectrum}, the required spectrum resource is still as large as 19 GHz \cite{UniEurope2016Identification}. On the other hand, energy constraint is another critical issue of IoT devices. Traditional transmitters in IoT devices use active radio frequency (RF) components such as converters and oscillators, which are costly and power-consuming, thus may not be suitable for low-power IoT devices. 
Therefore, novel spectrum and energy efficient communication technologies need to be developed for the future IoT.

One solution to achieve energy efficient IoT is ambient backscatter communication (AmBC) \cite{Liu2013}, in which a passive backscatter device (BD) modulates its information over ambient RF signals (e.g., cellular and WiFi signals) without requiring active RF components. While suitable for passive IoT, due to the spectrum sharing nature, the backscatter transmission in AmBC may suffer from severe direct-link interference (DLI) \cite{Wang2016Ambient}, resulting in performance degradation for the BD transmission. To tackle the DLI problem, in \cite{Guo2018Exploiting}, multiple receive antennas are used to suppress the DLI, while in \cite{Yang2017-p-}, a novel BD waveform is designed with clever interference cancellation by exploiting the cyclic prefix of ambient orthogonal frequency division multiplexing (OFDM) signal. In \cite{Zhang2018Constellation}, an improved Gaussian mixture model (GMM) based algorithm is proposed to recover the BD symbols by exploiting the received signal constellation information. In \cite{Yang-2017-p1-6}, a cooperative receiver with multiple antennas is designed to decode both the primary signal and the backscattered signal. In \cite{Darsena-2017-p1-1}, the capacity of AmBC system over ambient OFDM signals with perfect DLI cancellation is analyzed. In \cite{Kang-2017-p1-6}, the AmBC system is designed to maximize the ergodic capacity of the BD by jointly optimizing the ambient RF source's transmit power and the BD's reflection coefficient. In \cite{Duan2017Achievable}, the sum rate of the multiple-input multiple-output (MIMO) primary system and the multi-antenna BD transmission is analyzed. In \cite{Liu2018Backscatter}, the achievable rate region of the primary and BD transmissions is studied based on a new multiplicative multiple-access channel model. The above studies all assume certain forms of cooperation at the receiver side to cancel out the DLI or suppress the DLI effect.

In this paper, we propose a novel passive IoT transmission scheme, namely symbiotic radio (SR), in which a BD is integrated with a primary transmission. In the proposed SR, the primary transmitter (PT) is designed to support both the primary and BD transmissions, and the primary receiver (PR) needs to decode the information from the PT as well as the BD. Based on different transmission rate of the BD, the proposed SR can be further divided into parasitic SR (PSR), for which the BD transmission may introduce interferences to the primary transmission, and commensal SR (CSR), for which the two transmissions benefit from each other. One of the typical applications for SR is smart home, in which a smartphone recovers the data from both its serving WiFi AP and a served IoT sensor in its vicinity.

Compared with the conventional AmBC systems, in the proposed SR, the BD transmission shares not only the radio spectrum and RF source but also the receiver with the primary system. In this paper, we consider a basic SR model consisting of a multiple-input single-output (MISO) primary system and a single-antenna BD. The PT jointly designs its transmit beamforming to assist in the primary and BD transmissions, while the PR cooperatively decodes the signals from both the PT and the BD. Thus, the BD can realize opportunistic transmission with the aid of the primary system; on the other hand the achievable rate of the primary system can be improved by properly exploiting the additional signal path from the BD.

The main contributions of this paper are summarized as follows:
\begin{itemize}

\item First, we establish a general system model for the proposed SR and further investigate two practical setups, PSR and CSR, for which the symbol period for BD transmission is assumed to be either same as or much longer than that of the primary system, respectively.

\item Second, we analyze the achievable rates of the primary and BD transmissions under the two SR setups. Specifically, the achievable rate of the primary transmission is derived by treating the BD signal as an interference for PSR and a multipath signal component for CSR. After the PR cancels out the primary signal, the achievable rate of the BD transmission is obtained.

\item Third, transmit beamforming problems are formulated and solved, which aim to maximize the weighted sum rate of the primary and BD transmissions or to minimize the PT's transmit power under rate constraints.

\item Fourth, we propose a novel optimal beamforming structure to reduce the computational complexity. Specifically, the optimal transmit beamforming vector is shown to be a linear combination of the primary direct-link channel vector and the backscatter-link channel vector.

\item At last, numerical examples are presented to show that for the CSR system, the BD can realize its own transmission, and meanwhile enhance the primary transmission rate by providing an additional signal path for the primary system.

\end{itemize}


\begin{table}[t]\label{table:abbr}
\caption{List of abbreviations}
  \centering
    \begin{spacing}{1.2}{
    \begin{tabular}{|l|l|}
    \hline
    \textbf{Abbreviation} & \textbf{Description} \\
    \hline
      AmBC &     Ambient Backscatter Communication \\
    \hline
      BD &     Backscatter Device \\
    \hline
    CSCG & Circularly Symmetric Complex Gaussian\\
    \hline
    CSI & Channel State Information\\
    \hline
    CSR & Commensal Symbiotic Radio \\
    \hline
     DLI & Direct-link Interference \\
    \hline
    IoT & Internet of Things\\
    \hline
    MIMO  & Multiple-input Multiple-output   \\
    \hline
     MISO &  Multiple-input Single-output \\
    \hline
    OFDM & Orthogonal Frequency Division \\
    &Multiplexing\\
    \hline
    PDF & Probability Density Function\\
    \hline
    PR & Primary Receiver\\
    \hline
    PSD & Positive Semi-definite\\
    \hline
    PSR & Parasitic Symbiotic Radio\\
    \hline
    PT & Primary Transmitter\\
    \hline
    RF & Radio Frequency\\
    \hline
    SDR & Semi-definite Relaxation\\
    \hline
    SIC & Successive Interference Cancellation\\
    \hline
    SINR & Signal to Interference plus Noise Ratio\\
    \hline
    SNR & Signal to Noise Ratio\\
    \hline
    SR & Symbiotic Radio\\
    \hline
    TPM & Transmit Power Minimization\\
    \hline
    WSRM &  Weighted Sum-rate Maximization\\
    \hline
    \end{tabular}
  }
  \end{spacing}
\end{table}

The rest of this paper is organized as follows. In Section \ref{sec:system-model}, we present the SR system model. In Section \ref{sec:ARA_analysis}, we derive the achievable rates of the primary and the BD transmissions for both PSR and CSR. In Section \ref{sec:ProblemF}, we formulate the weighted sum-rate maximization problem and the transmit power minimization problem. In Section \ref{sec:solution}, we present the SDR-based solutions to the formulated problems. In Section \ref{sec:LowC}, a more efficient algorithm with lower complexity is presented based on a novel beamforming structure. In Section \ref{sec:NumR}, numerical results are presented for performance evaluations. Finally, the paper is concluded in Section \ref{sec:Conclusion}.

The main notations in this paper are listed as follows: The lowercase, boldface lowercase, and boldface uppercase letters such as  $t$, $\mathbf{t}$, and $\mathbf{T}$ denote the scalar, vector, and matrix, respectively. $|t|$ denotes the absolute value of $t$. $\|\mathbf{t}\|$ denotes the norm of vector $\mathbf{t}$. ${\cal{CN}}(\mu, \sigma^2)$ denotes the circularly symmetric complex Gaussian (CSCG) distribution with mean $\mu$ and variance $\sigma^2$. $\mathbb{E}[\cdot]$ denotes the statistical expectation. $t^{\ast}$ denotes the conjugate of $t$. $\mathbf{T}^{\mathrm{T}}$ and $\mathbf{T}^{\mathrm{H}}$ denotes the transpose and conjugate transpose of matrix $\mathbf{T}$, respectively. Finally, the  list of abbreviations commonly appeared in this paper is given in Table \ref{table:abbr}.


\section{System Model}
\label{sec:system-model}

\begin{figure}[t]
    \centering\includegraphics[width=.63\columnwidth]{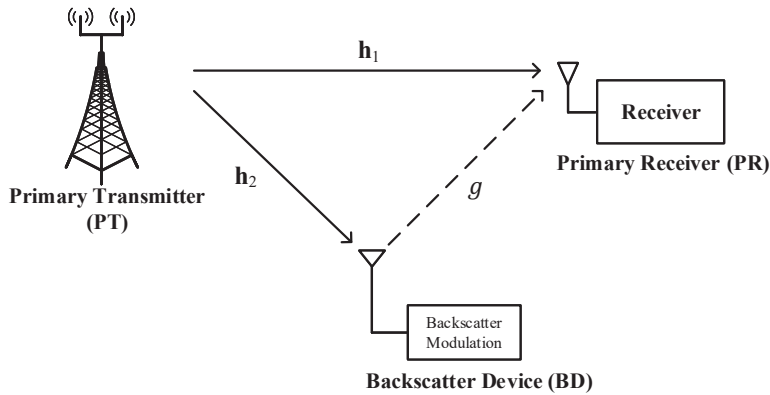} %
    \caption{System model of a symbiotic radio.}\label{fig:systemmodel}
\end{figure}

Fig.~\ref{fig:systemmodel} shows the system model of a symbiotic radio (SR) consisting of three nodes, namely the primary transmitter (PT) equipped with $M$ ($M > 1$) antennas, the single-antenna primary receiver (PR) and the single-antenna backscatter device (BD). The PT performs multi-antenna beamforming to transmit its primary information to the PR, and at the same time enables the BD to transmit information to the PR. Specifically, the BD modulates its own information over the incident (primary) signal from the PT by intelligently varying its reflection coefficient. The SR thus shares not only the same spectrum but also the same receiver with the primary system.

Block flat-fading channel models are considered in this paper. During each fading block, the direct-link channel from PT to PR is denoted by $\mathbf{h}_1 = [h_{1,1}, \ldots, h_{M,1}]^\mathrm{T} \in\mathbb C^{M\times 1}$, where $h_{m,1}, \forall m$, denotes the the channel coefficient between the PT's $m$-th antenna and the PR's antenna. Meanwhile, the backscatter-link channel, denoted by $g\mathbf{h}_{2}$, is the multiplication of the forward-link channel from PT to BD, denoted by $\mathbf{h}_{2}= [h_{1,2}, \ldots, h_{M,2}]^\mathrm{T} \in \mathbb{C}^{M\times 1}$, and the backward-link channel from BD to PR, denoted by $g \in \mathbb{C}$. We assume that the SR system operates in the time-division-multiplexing (TDD) mode, and the PT and the PR have perfect channel state information (CSI) of the direct link and the backscatter link. In practice, the CSI can be obtained by the training-based channel estimation scheme with two steps. First, the BD switches its impedance into the matched state, and the PT estimates the direct-link channel $\mathbf{h}_1$ via channel reciprocity. Second, the BD switches its impedance into a fixed and known backscatter state, and the PT estimates the backscatter-link channel $g \mathbf{h}_2$ by subtracting the estimated direct-link channel component ${\mathbf{h}}_1$ from the estimated composite channel $\mathbf{h}_1 + g \mathbf{h}_2$.


\section{Achievable Rate Analysis}\label{sec:ARA_analysis}
In this section, we analyze the achievable rate performance of the proposed SR. Let $s(n)$ be the signal transmitted by the PT with symbol period $T_s$, and $s(n)$ is assumed to follow the standard CSCG distribution, i.e., $s(n) \sim \mathcal{CN}(0,1)$. Denote the beamforming vector of the PT by $\mathbf{w}\in\mathbb{C}^{M\times 1}$. Let $c(n)$ be the BD signal to be transmitted, with symbol period $T_c$. The $c(n)$ varies with different reflection coefficients and is assumed to be distributed\footnote{We assume that the Gaussian codewords herein to derive the maximum achievable rate of the SR.} as $\mathcal{CN}(0,1)$. The backscattered signal from the BD is thus $\sqrt{\alpha} c(n)$, where the power reflection coefficient $\alpha \in[0,1]$ controls the power of the backscattered signal by the BD. It is noticed that there is no additive noise in the BD, since its integrated circuit only includes passive components \cite{Fuschini-2008-p33-35,Qian2017Noncoherent}.


In the following, we consider two setups based on different relationships between $T_s$ and $T_c$. One is PSR for which $T_s = T_c$, and the other is CSR, for which $T_c = N T_s$, where $N$ is an integer and $N \gg 1$.



\subsection{PSR Setup}\label{sec:setupESP}
Let $p$ be the transmit power of the PT. The PR receives the backscattered signal from the BD as well as the primary signal transmitted from the PT. In the $n$-th symbol period, the received signal at the PR, denoted by $y(n)$, is thus given by
\begin{equation}\label{eq:Tr1}
y(n) = \sqrt {p} \mathbf{h}_1^\mathrm{ H} \mathbf{w} s(n) + \sqrt{p} \sqrt{\alpha} c (n)g\mathbf{h}_2^\mathrm{H} \mathbf{w}s(n)  + z(n),
\end{equation}
where $z(n)$ is the additive white Gaussian noise (AWGN) with zero mean and power $\sigma^2$, i.e., $z(n)\sim{~}\mathcal{CN}(0,\sigma^2)$.


In practice, the direct-link signal is typically stronger\footnote{Since the analog-to-digital convertor (ADC) in the receiver often has large dynamic range (e.g., 49.9 dB for an 8-bit ideal ADC\cite{Walden-1999-p539-550}.) and the line-of-sight (LoS) pathloss due to the transmission from the BD to PR is usually within this range (e.g., 28 dB for 5m distance), the two received signals generally will not exceed the dynamic range of ADC.} than the backscatter-link signal, due to the following two facts. First, the backscatter-link channel suffers from double attenuations, i.e., the forward-link channel $\mathbf{h}_2$ and the backward-link channel $g$. Second, compared to the incident primary signal, the backscattered signal from the BD further suffers from an obvious power loss due to the backscattering operation. As a result, the PR can first decode the primary signal $s(n)$, then cancels out the decoded signal $\hat{s}(n)$ from its received signal, and finally detects the BD signal $c(n)$. In the following, we analyze the achievable rate performance of such decoding scheme.



Since $s(n)$ and $c(n)$ have the same symbol rate, when the PR decodes the primary signal $s(n)$, it treats the BD signal as the background noise of which the average power is
$\mathbb{E}\left[\alpha p|g|^2 |c(n)|^2 |\mathbf{h}^\mathrm{ H}_2\mathbf{w}|^2\right]=\alpha p|g|^2 |\mathbf{h}^\mathrm{ H}_2\mathbf{w} |^2$. Thus the signal-to-interference-plus-noise ratio (SINR) for decoding $s(n)$ at the PR is given by
\begin{equation}\label{eq:SINR_PR}
    \gamma_{s}^{(1)}=\frac{{p | {\mathbf{h}_1^\mathrm{H}\mathbf{w}}|^2}}{{\alpha p|g|^2 | {\mathbf{h}_2^\mathrm{H}\mathbf{w}}  |^2 + {\sigma ^2}}}.
\end{equation}
The corresponding data rate of the primary system can be written as
\begin{equation}\label{eq:Rs1}
    {R_{s}^{(1)}} =  {{{\log }_2}(1 + \gamma_{s}^{(1)})}.
\end{equation}

After obtaining an estimation of the primary signal $\hat{s}(n)$, the PR utilizes the successive-interference-cancellation (SIC) technique to decode the BD signal $c(n)$. That is, the received primary signal component $\sqrt{p}\mathbf{h}_1^\mathrm{ H}\mathbf{w}\hat{{s}}(n)$ is subtracted from the received signal $y(n)$, yielding the following intermediate signal
\begin{equation}\label{eq:yc_est}
    \hat{y}_c(n)=y(n)-\sqrt {p} \mathbf{h}_1^\mathrm{ H}\mathbf{w}\hat{s}(n).
\end{equation}
Assuming that the primary signal is removed perfectly, we have
\begin{equation}\label{eq:yc}
    \hat{y}_c(n)=\sqrt{\alpha}\sqrt{p} s(n) g\mathbf{h}_2^\mathrm{H} \mathbf{w} c(n) + z(n).
\end{equation}
Given the primary signal $s(n)$, the signal-to-noise ratio (SNR) for decoding the BD signal is written as
\begin{equation}\label{eq:SNR_SR}
    \gamma_{c}^{(1)} (s) = \frac{{\alpha p |s(n)|^2 |g|^2| { \mathbf{h}_2^\mathrm{ H}\mathbf{w}} |^2}}{{{\sigma ^2}}}.
\end{equation}
Thus the average data rate of the BD transmission is written by
\begin{equation}\label{eq:Rc1}
    { R_{c}^{(1)}} = \mathbb{E}_s\left[ {{{\log }_2} \left( 1 + \gamma_{c}^{(1)} (s) \right)} \right].
\end{equation}

When decoding $c(n)$, the primary signal $s(n)$ plays the role of fast-varying channel responses. The squared envelope $|s(n)|^2$ of $s(n)$ follows an exponential distribution, and its probability density function (PDF) is $f(x)=e^{-x},x>0$. Thus, the BD (i.e., backscatter-link) transmission rate $R_{c}^{(1)}$ can be derived as follows
\begin{align}\label{eq:Rc_pdf}
  { R_{c}^{(1)}} &= \int_{\rm{0}}^{{\rm{ + }}\infty } {{\mathrm{e}^{ - x}}{{\log }_2}} (1 + \beta x)\mathrm{d}x \nonumber\\ &={{-\mathrm{e}}^{\frac{1}{\beta }}}\rm{Ei}\left( - \frac{1}{\beta } \right)log_2\mathrm{e},
\end{align}
where $\beta = \frac{{{{\alpha  p|g|^2 \left| {\mathbf{h}_2^\mathrm{H}\mathbf{w}} \right|}^2}}}{{{\sigma ^2}}}$ is the average received SNR of the backscatter link, and $\mathrm{Ei} {\left( x \right)} \triangleq \int_{-\infty}^{x} \frac{1}{u} \mathrm{e}^{u} \mathrm{d}u$ is defined for the exponential integral.

\remark
The ${{-\mathrm{e}}^{\frac{1}{x }}}\mathrm{Ei} \left( - \frac{1}{\it{x} }\right)$ is a monotonically increasing and concave function of $x$, for $x\geq0$. This can be easily verified by its first and second derivatives.

\subsection{CSR Setup}\label{sec:setupUSP}
In this subsection, we thus consider the CSR setup in which $T_c= NT_s$, where $N$ is an integer, and $N \gg 1$. Compared with the PSR setup, the BD transmission in CSR has much low rate than the primary transmission, thus it can provide an additional signal component by its scattering.

To differentiate CSR from the PSR, we let $c$ be the BD signal to be transmitted in one particular BD symbol period, which covers $N$ primary symbol periods. Thus, in the $n$-th primary symbol period within one BD symbol period, for $n=1, \ldots, N$, the received signal at the PR is given by
\begin{equation}\label{eq:Tr2}
y(n) = \sqrt {p} \mathbf{h}_1^\mathrm{ H} \mathbf{w} s(n) + \sqrt{p} \sqrt{\alpha} c g\mathbf{h}_2^\mathrm{H} \mathbf{w}s(n)  + z(n).
\end{equation}
The second signal term in~\eqref{eq:Tr2} can be viewed as the output of the primary signal $s(n)$ passing through a slowly varying channel $\sqrt{\alpha} c g\mathbf{h}_2$. Thus the PR first decodes the primary signal $s(n)$ by treating the BD signal as a multipath component. The equivalent channel for decoding $s(n)$ is denoted by $\mathbf{h}_\mathrm{eq}=\mathbf{h}_1 +\sqrt{\alpha}{c} g\mathbf{h}_2$. Since the PR has no prior knowledge about the BD signal $c$, a training symbol from the PT is required to estimate the equivalent channel  $\mathbf{h}_\mathrm{eq}$. Given $c$, the SNR for decoding $s(n)$ is written as
\begin{equation}\label{eq:SNR_PR2}
    \gamma_{s}^{(2)}(c) = \frac{p \left| \mathbf{h}_\mathrm{eq}^\mathrm{H}(c)  \mathbf{w} \right|^2}{\sigma ^2}.
\end{equation}

With a given $c$, the achievable rate of the direct link is thus given by
\begin{equation}\label{eq:NoncoherentDetection}
\tilde{R}_{s}^{(2)}(c)=\log_2\left(1+\gamma_{s}^{(2)}(c)\right),
\end{equation}
where we have ignored the training overhead in each BD symbol period due to large $N$.

Thus, for sufficiently large $N$, the average primary rate is
\begin{equation}\label{eq:Rs2}
    {R_{s}^{(2)}} = \mathbb{E}_c\left[ {{{\log }_2}(1 + \gamma_{s}^{(2)}(c))} \right],
\end{equation}
where the expectation is taken over the random variable $c$.

\proposition
$\gamma_{s}^{(2)}$ is distributed as a noncentral chi-square distribution $\chi^2$ with the freedom of $2$, the non-centrality parameter $\lambda = \frac{p \left|\mathbf{h}^\mathrm{H}_1 \mathbf{w}\right|^2 }{\sigma ^2}$ and the Gaussian variance parameter $\Sigma= \frac{p \alpha |g|^2 \left|\mathbf{h}^\mathrm{H}_2 \mathbf{w}\right|^2 }{2\sigma ^2}$. Its PDF is given by
\begin{equation}\label{eq:NoncenChipdf}
f(x)= \frac{1}{2\Sigma}\mathrm{e}^{\left(-\frac{x+\lambda}{2\Sigma}\right)}I_0\left(\frac{\sqrt{x\lambda}}{\Sigma}\right),
\end{equation}
where $I_0\left(\cdot\right)$ is a modified Bessel function of the first kind given by
\begin{equation}\label{eq:Bessel}
  I_0(x) = \sum_{m=0}^{\infty}\frac{1}{m!\Gamma(m+1)}\left(\frac{x}{2}\right)^{2m}.
\end{equation}

\begin{IEEEproof}
Please refer to Appendix \ref{proof:pro_Nonchi}.
\end{IEEEproof}

Notice that the non-centrality parameter $\lambda$ can be explained as the SNR of the direct link, while the Gaussian variance related parameter $2\Sigma$ can be interpreted as the SNR of the backscatter link. Let $x=\gamma_{s}^{(2)}$. From~\eqref{eq:NoncenChipdf}, the achievable rate $ R_{s}^{(2)}$ in~\eqref{eq:Rs2} can be expanded as follows,
\begin{equation}\label{eq:Rs2_integral}
{ R_{s}^{(2)}} = \int_{0}^{+\infty} \log_2(1+x) f(x) \mathrm{d} x.
\end{equation}

In order to obtain analytical insights, we consider the asymptotic case with high SNR $\gamma_{s}^{(2)}$.
\proposition For the case of SNR $\gamma_{s}^{(2)} \rightarrow + \infty$, the primary rate $R_{s}^{(2)}$ can be obtained with a closed-form as follows
\begin{equation}\label{eq:Rs2_closed}
R_{s}^{(2)} = \log_2 \lambda- \mathrm{Ei}\left(-\frac{\lambda}{2\Sigma}\right) \log_2 \mathrm{e}.
\end{equation}
\begin{IEEEproof}
Please refer to Appendix \ref{proof:pro_Rs}.
\end{IEEEproof}

Clearly, the first term in \eqref{eq:Rs2_closed} of Proposition 2 can be interpreted as the achievable rate for a traditional MISO system with transmit beamforming. Moreover, we have the following important observations for Proposition 2.
\remark First, compared to the traditional MISO system, the primary transmission in the SR achieves a rate gain of $\Delta R_{s}^{(2)} = - \mathrm{Ei}\left(-\frac{\lambda}{2\Sigma}\right) \log_2 \mathrm{e}$, since $\Delta R_{s}^{(2)}>0$. This implies that the existence of the backscattering BD can enhance the primary transmission rate by providing an additional scattered path for the primary system. Second, the rate gain of the primary system $\Delta R_{s}^{(2)}$ increases as the backscatter-link SNR $2\Sigma$ increases, for any given direct-link SNR $\lambda$, since the exponential integer function $\mathrm{Ei}(x)$ is monotonically decreasing for $x < 0$.





After decoding $s(n)$, the PR also applies the SIC technique to remove the direct-link interference. In a BD symbol duration, we denote the primary signal vector by $\mathbf{s}=\left[s(1),s(2),\dots,s(N)\right]^\mathrm{T}$, the noise vector by $\mathbf{z}=\left[z(1),z(2),\dots,z(N)\right]^\mathrm{T}$ and the received signal vector after the interference cancellation by $\mathbf{\hat{y}_c}=\left[\hat{y}_c(1),\hat{y}_c(2),\dots,\hat{y}_c(N)\right]^\mathrm{T}$. Assuming that the primary signal component is removed perfectly, we obtain the intermediate signal in a vector form as
\begin{equation}\label{eq:yc2}
    \mathbf{\hat{y}_c}=\sqrt{\alpha} \sqrt{p}{g}\mathbf{h}_2^\mathrm{H} \mathbf{w} \mathbf{s} c + \mathbf{z}.
\end{equation}

Since $\mathbb{E} [|s(n)|^2]=1$, the SNR for decoding BD symbol $c$ via maximal ratio combining (MRC) can be approximated as (assuming $N \gg 1$)
\begin{equation}\label{eq:SNR_SR2}
    \gamma_{c}^{(2)} = \frac{N \alpha  p |g|^2\left| {\mathbf{h}_2^\mathrm{ H}\mathbf{w}} \right|^2}{\sigma^2}.
\end{equation}

In the CSR setup, since only one BD symbol is transmitted during $N$ successive primary-symbol periods, the primary signal $s(n)$ can be viewed as a spread-spectrum code with length $N$ for BD symbols. Accordingly, the SNR for decoding BD symbol $\gamma_{c}^{(2)}$ is increased by $N$ times, at the cost of symbol rate decreased by $\frac{1}{N}$. Hence, the BD achievable rate is given by 
\begin{equation} \label{eq:Rc2}
    {R_{c}^{(2)}} = \frac{1}{N} {{{\log }_2}(1 + \gamma_{c}^{(2)})}.
\end{equation}

\section{Transmit Beamforming Problem Formulation}\label{sec:ProblemF}
In this section, to further investigate the performance of the proposed SR, we consider two transmit beamforming optimization problems, i.e.,  weighted sum-rate maximization (WSRM) problem and transmit power minimization (TPM) problem.

\subsection{Weighted Sum-Rate Maximization}
\label{sec:WSRMP}
In this subsection, we aim to maximize the weighted sum of the primary rate and the BD rate by optimizing the transmit beamforming vector $\mathbf{w}$. A general WSRM problem can be formulated as follows
\begin{subequations}
\begin{eqnarray}\label{eq:WSRMP_General}
\max_{\mathbf{w}}&& \rho R_s^{(i)} + (1-\rho) R_c^{(i)},  \\
\mathrm{s.t.} &&\|\mathbf{w}\|^2=1,\label{eq:PGen}
\end{eqnarray}
\end{subequations}
where the weight factor $\rho\in[0,1]$, the index $i\in\{1,2\}$ indicates the PSR setup and the CSR setup, respectively, and \eqref{eq:PGen} is the normalization constraint for the transmit beamforming design.

By following \cite{Shang2011Multiuser}, the achievable rate region can be adopted to characterize the optimal rate tradeoff between the primary and BD transmissions. Specifically, the rate region consists of all the achievable rate pairs that can be achieved by the PT and BD transmissions under the considered beamforming scheme. By varying the weight factor $\rho$ in \eqref{eq:WSRMP_General}, a sequence of WSRM problems can be solved to obtain the Pareto boundary for the rate region of the SR with transmit beamforming.




For the PSR setup, from \eqref{eq:Rs1} and \eqref{eq:Rc_pdf}, we have the following WSRM problem

\begin{subequations}
\textbf{\underline{P1:}}\label{eq:P1}
\begin{align}
\max_{\mathbf{w}}~~& R_1 (\mathbf{w}) \triangleq { \rho \log_2\left(1+\frac{{p | {\mathbf{h}_1^\mathrm{H}\mathbf{w}}|^2}}{{\alpha p|g|^2 | {\mathbf{h}_2^\mathrm{H}\mathbf{w}}  |^2 + {\sigma ^2}}}\right)} - \nonumber\\
~&(1 \!-\! \rho){\mathrm{e}^{\frac{\sigma^2}{\alpha  p|g|^2\left|\mathbf{h}_2^\mathrm{H} \mathbf{w}\right|^2 }}}{\mathrm{Ei}}\left( \! - \frac{\sigma^2}{\alpha  p|g|^2\left|\mathbf{h}_2^\mathrm{H} \mathbf{w}\right|^2 } \! \right)\log_2\mathrm{e} \\
\mathrm{s.t.}~~
&\|\mathbf{w}\|^2=1.\label{eq:P1C1}
\end{align}
\end{subequations}

For the CSR setup, from \eqref{eq:Rs2} and \eqref{eq:Rc2}, we have the following WSRM problem

\textbf{\underline{P2:}}\label{eq:P2}
\begin{subequations}
\begin{align}
\max_{\mathbf{w}}~~& R_2 (\mathbf{w}) \triangleq \frac{1 \! -\! \rho}{N} {{{\log }_2}\left(1 \!+\! \frac{N \alpha p|g|^2 \left| {\mathbf{h}_2^\mathrm{ H}\mathbf{w}} \right|^2}{\sigma^2}\right)} \!+\! \nonumber\\
~&\rho \mathbb{E}_c\left[ {{{\log }_2}\left(1 + \frac{{p\left|  {(\mathbf{h}_1 + \sqrt{\alpha}{c}g\mathbf{h}_2)^\mathrm{H}\mathbf{w}} \right|^2}}{{{\sigma ^2}}} \right)} \right] \\
\mathrm{s.t.}~~&\|\mathbf{w}\|^2=1. \label{eq:P2C1}
\end{align}
\end{subequations}

Both (P1) and (P2) are non-convex optimization problems, and thus it is difficult to obtain their optimal solutions in general.


\subsection{Transmit Power Minimization}
\label{sec:PMP}
Since energy consumption is another important performance metric, in this subsection, we aim to minimize the PT's transmit power under given primary and BD rate requirements by optimizing the transmit beamforming vector $\mathbf{w}$ and the transmit power $p$ jointly. A general optimization problem is given by
\begin{subequations}
 \begin{align}
  \min_{\mathbf{w},p} & \quad\quad p  \\
  \mathrm{s.t.} & \quad\quad R_s^{(i)} \geq \epsilon_s, \\
  &\quad\quad R_c^{(i)} \geq \epsilon_c,  \\
  &\quad\quad \left\|\mathbf{w}\right\|^2=1 ,
\end{align}
\end{subequations}
where $\epsilon_s$ and $\epsilon_c$ are the rate requirements of the primary system and the BD, respectively.

For the PSR setup, the rate requirements can be equivalently converted into the SINR/SNR constraints. Then, the TPM problem can be rewritten as

\textbf{\underline{P3:}}
\begin{subequations}
\begin{align}
  \min_{\mathbf{w},p} & \quad\quad p\\
  \mathrm{s.t.} & \quad\quad \frac{{p | {\mathbf{h}_1^\mathrm{H}\mathbf{w}}|^2}}{{\alpha  p|g|^2 | {\mathbf{h}_2^\mathrm{H}\mathbf{w}}  |^2 + {\sigma ^2}}} \geq 2^{\epsilon_{s}}-1,  \\
  &\quad\quad \frac{{{{\alpha p|g|^2 \left| {\mathbf{h}_2^\mathrm{H}\mathbf{w}} \right|}^2}}}{{{\sigma ^2}}} \geq \gamma_\beta(\epsilon_c), \\
  &\quad\quad \left\|\mathbf{w}\right\|^2=1,
\end{align}
\end{subequations}
where $\gamma_\beta(\epsilon_c)$ is the root of the equation $R_c^{(1)}=\epsilon_c$, that is
\begin{equation}\label{eq:Epsilonc2SNR}
  {{-\mathrm{e}}^{\frac{1}{\beta }}}\rm{Ei}\left( - \frac{1}{\beta } \right)log_2\mathrm{e}=\epsilon_c.
\end{equation}


By converting the BD rate requirement into the SNR constraint, the TPM problem can be rewritten as follows

\textbf{\underline{P4:}}
\begin{subequations}
\begin{align}
  \min_{\mathbf{w},p} & \quad\quad p  \\
  \mathrm{s.t.} & \quad \mathbb{E}_c\left[ {{{\log }_2}\left(1 + \frac{{p\left|  {(\mathbf{h}_1 + \sqrt{\alpha} {c}g\mathbf{h}_2)^\mathrm{H}\mathbf{w}} \right|^2}}{{{\sigma ^2}}} \right)} \right] \geq \epsilon_{s},\label{eq:P4-C1}  \\
  &\quad \frac{{{{\alpha p|g|^2 \left| {\mathbf{h}_2^\mathrm{H}\mathbf{w}} \right|}^2}}}{{{\sigma ^2}}} \geq \frac{2^{N \epsilon_c}-1}{N},  \\
  &\quad \left\|\mathbf{w}\right\|^2=1.
\end{align}
\end{subequations}
However, \eqref{eq:P4-C1} cannot be converted into a SNR constraint in (P4), since it is difficult to obtain a closed-form expression for the primary rate in terms of SNR.

It is easy to see that the TPM problems are always feasible, provided of course that none of the channel vectors is identically zero and the channel vectors $\mathbf{h}_1$ and $\mathbf{h}_2$ are not parallel to each other. However, it is also verified that both (P3) and (P4) are non-convex optimization problems which are difficult to solve optimally.
%
%

\section{Proposed Solutions}\label{sec:solution}
In this section, we propose algorithms to obtain generally suboptimal solutions to the problems formulated in the previous section.


\subsection{Weighted Sum-Rate Maximization}
\subsubsection{PSR Setup} Denote $\mathbf{v}=\sqrt{p}\mathbf{w}$, $\mathbf{H}_{1}=\mathbf{h}_{1}\mathbf{h}_{1}^\mathrm{ H}$ and $\mathbf{H}_{2}=\alpha |g|^2 \mathbf{h}_{2}\mathbf{h}_{2}^\mathrm{ H}$ for convenience. By introducing the new variable $\mathbf{W}=\mathbf{vv}^{\mathrm{H}}$, (P1) is recast as the following equivalent optimization problem with a positive semi-definite (PSD) matrix variable $\mathbf{W}$.

\begin{algorithm}[t]
\caption{ for solving (P1-SDR)}\label{algorithm1}
\begin{algorithmic}[1]
\REQUIRE
\textcolor{black}{The power reflection coefficient $\alpha$; transmit power $p$; the CSI $\mathbf{h}_1$, $g \mathbf{h}_2$ and the noise power $\sigma^2$.}
\ENSURE
\textcolor{black}{The solution for (P1-SDR) $\mathbf{W}^{\star}$.}
\STATE Initialization: $\xi = \sigma^2$, the interval $\Delta\xi$, and iteration index $k=1$.
\WHILE {$\xi \leq \alpha p|g|^2 \left\|\mathbf{h}_2\right\|^2 + \sigma^2 $}
\STATE Given $\xi$, solve {\rm{(P1-SDR)}} by using CVX to obtain the optimal $\mathbf{W}_k^{\star} (\xi)$ and objective value $C_k (\xi)$.
\STATE $\xi \leftarrow \xi + \Delta\xi$.
\STATE $k \leftarrow k+1$.
\ENDWHILE
\STATE Obtain the optimal solution to (P1-SDR) as $\mathbf{W}^{\star} = \mathbf{W}_{k^{\star}}$, where $k^{\star} = \arg \underset{k}{\max} \;\; C_k(\xi)$.
\end{algorithmic}
\end{algorithm}

\textbf{\underline{P1-PSD:}}
\begin{subequations}
\begin{eqnarray}\label{eq:P1-PSD}
\max_{\mathbf{W}}&&  \rho{{{\log }_2}\left(1 + \frac{\Tr(\mathbf{H}_{1}\mathbf{W})}{\Tr(\mathbf{H}_{2}\mathbf{W})+\sigma^2}\right)} - \hspace{-0.1cm}(1\hspace{-0.1cm}-\hspace{-0.1cm}\rho)\hspace{-0.3cm}\quad {\mathrm{e}^{\frac{\sigma^2}{\Tr(\mathbf{H}_2 \mathbf{W}) }}}\rm{Ei}\left( - \frac{\sigma^2}{\Tr(\mathbf{H}_2 \mathbf{W}) } \right)log_2\mathrm{e} \label{eq:P1-PSDObj1} \\
\mathrm{s.t.} && \Tr({\mathbf{W}})=p, \label{eq:P1-PSDC1} \\
&&\mathrm{Rank}(\mathbf{W}) = 1.\label{eq:P1-PSDC3}
\end{eqnarray}
\end{subequations}

To solve this problem, we replace the denominator in the objective function~\eqref{eq:P1-PSDObj1} with an auxiliary  variable $\xi \triangleq \Tr(\mathbf{H}_2\mathbf{W})+\sigma^2$ and add an equality constraint $\Tr(\mathbf{H}_2\mathbf{W})+\sigma^2 = \xi$ accordingly. Moreover, by relaxing the nonconvex rank-one constraint \eqref{eq:P1-PSDC3}, (P1-PSD) can be recast to the following semi-definite relaxation (SDR) problem \cite{CVXBoyd04}

\textbf{\underline{P1-SDR:}}
\begin{subequations}
\begin{eqnarray}\label{P1-SDR}
\max_{\mathbf{W},\xi}&& {{{\rho \log }_2}\left(1+\frac{\Tr(\mathbf{H}_1\mathbf{W})}{\xi} \right)} -  \ (1-\rho) {\mathrm{e}^{\frac{\sigma^2}{\xi-\sigma^2 }}}\rm{Ei}\left( - {\frac{\sigma^2}{\xi-\sigma^2 }} \right)log_2\mathrm{e}\\
\mathrm{s.t.}&&\Tr(\mathbf{W})= p,\\
&&\Tr(\mathbf{H}_2\mathbf{W})+\sigma^2= \xi.
\end{eqnarray}
\end{subequations}
Notice that for a given $\xi$, (P1-SDR) is a convex optimization problem which can be solved optimally by using software tools such as CVX \cite{grant2008cvx}. Then the optimal $\xi^{\star}$ can be obtained by one-dimensional exhaustive search over $\xi$. The details for solving (P1-SDR) are summarized in Algorithm 1.

If the SDR solution $\mathbf{W}^{\star}$ obtained by Algorithm 1 is of rank one, i.e., $\mathbf{W}^{\star}=\mathbf{w}^{\star}(\mathbf{w}^{\star})^{\mathrm{H}}$, then $\frac{\mathbf{w}^{\star}}{\sqrt{p}}$ is the optimal solution to (P1). Otherwise, we use the randomization-based method \cite{sidiropoulos2006transmit} to obtain an approximate (suboptimal) solution to (P1). Based on $\mathbf{W}^{\star}$, the steps to find the solution to (P1) are summarized in Algorithm~\ref{Algorithm_rand}.

\subsubsection{CSR Setup}
Let $\mathbf{W}=\mathbf{vv}^{\mathrm{H}}$, $\mathbf{H}_\mathrm{eq}=\mathbf{h}_\mathrm{eq}\mathbf{h}_\mathrm{eq}^\mathrm{H}$ and $\mathbf{H}_{2}=\alpha |g|^2 \mathbf{h}_{2}\mathbf{h}_{2}^\mathrm{ H}$ for convenience. Then (P2) is recast into the following equivalent problem

\begin{algorithm}[t!]
\caption{ for solving (P1)} \label{Algorithm_rand}
\begin{algorithmic}[1]
\REQUIRE \textcolor{black}{The solution to (P1-SDR) $\mathbf{W}^{\star}$.}
\ENSURE \textcolor{black}{The beamforming solution $\mathbf{w}^{\star}$.}
\STATE Initialization: the solution $\mathbf{W}^{\star}$ to (P1-SDR), a large positive integer $D$. \\
\STATE Compute the singular value decomposition (SVD) of $\mathbf{W}^{\star}$ as $\mathbf{W}^{\star}=\mathbf{U{\Sigma}U}^\mathrm{ H}$, with $\mathbf{U}=[\mathbf{u}_1 \cdots \mathbf{u}_M]$. \\
\IF {$\Rank ({\mathbf{W}^{\star}})=1$, }
\STATE $\mathbf{w}^{\star} = \mathbf{u}_1$.
\ELSE
\FOR{$d=1,\ldots,D$}
\STATE Generate a random vector $\mathbf{v}_d =\mathbf{U} \mathbf{\Sigma}^{\frac{1}{2}} \mathbf{e}_d$, where $\mathbf{e}_d=[\mathrm{e}^{j{\theta}_1},\mathrm{e}^{j{\theta}_2},...,\mathrm{e}^{j{\theta}_n}]^\mathrm{ H}$ and $\theta_i$ follows the uniform distribution $U(0,2\pi)$.
\ENDFOR
\RETURN $\mathbf{w}^{\star}=\frac{\mathbf{v}^{\star}}{\sqrt{p}}$, where $\mathbf{v}^{\star}= \arg \underset{d \in \calD} {\max} \ \ R_1 (\mathbf{v}_d)$.
\ENDIF
\end{algorithmic}
\end{algorithm}
\textbf{\underline{P2-PSD:}}
\begin{subequations}
\begin{eqnarray}\label{eq:P2-PSD}
\max_{\mathbf{W}}&& \rho\mathbb{E}_c\left[ {{{\log }_2}\left(1 + \frac{\Tr(\mathbf{H}_\mathrm{eq}(c)\mathbf{W})}{\sigma^2}\right)} \right]+  (1-\rho)\frac{1}{N} {{{\log }_2}\left(1 + \frac{N \Tr(\mathbf{H}_{2}\mathbf{W})}{\sigma^2}\right)} \label{P2Sobj}\\
\mathrm{s.t.}&&\Tr(\mathbf{W})=p,\label{P2SC1}\\
&&\mathrm{Rank}(\mathbf{W})=1.\label{P2SC2}
\end{eqnarray}
\end{subequations}
Similar to (P1-PSD), the SDR of (P2-PSD) is a convex optimization problem, which can be solved optimally and efficiently. Once obtaining the SDR solution $\mathbf{W}^{\star}$ to (P2-PSD), we can use an algorithm similar to Algorithm~\ref{Algorithm_rand} to find a generally approximate solution $\mathbf{w}^{\star}$ to (P2). The details are omitted here for brevity.

\subsection{Transmit Power Minimization}
A similar variable transformation as the WSRM problem can be applied to the TPM problem. The optimization problem (P3) is thus recast into the following equivalent problem

\textbf{\underline{P3-PSD:}}
\begin{subequations}
\begin{align}
  \min_{\mathbf{W}} & \quad\quad \mathrm{Tr}(\mathbf{W})  \\
 \mathrm{s.t.}&\quad\quad \frac{\Tr(\mathbf{H}_{1}\mathbf{W})}{\Tr(\mathbf{H}_{2}\mathbf{W})+\sigma^2} \geq 2^{\epsilon_s}-1, \\
  &\quad\quad \frac{\Tr(\mathbf{H}_{2}\mathbf{W})}{\sigma^2} \geq \gamma_\beta(\epsilon_c),\\
  &\quad\quad \mathrm{Rank}(\mathbf{W})=1.
\end{align}
\end{subequations}
Without the rank-one constraint, the SDR problem of (P3-PSD) can be solved by the CVX. Based on the SDR solution, Algorithm~\ref{Algorithm_TPM} is designed to find a rank-one solution $\mathbf{w}^\star$ together with a transmit power $p^\star$ to (P3).

\begin{algorithm}[t!]
\caption{ for solving (P3)} \label{Algorithm_TPM}
\begin{algorithmic}[1]
\REQUIRE \textcolor{black}{The SDR solution to (P3-PSD) $\mathbf{W}^{\star}$.}
\ENSURE \textcolor{black}{The transmit power $p^\star$ and the beamforming solution $\mathbf{w}^{\star}$.}
\STATE Initialization: the SDR solution $\mathbf{W}^{\star}$ to (P3-PSD), a large positive integer $D$. \\
\STATE Compute the singular value decomposition (SVD) of $\mathbf{W}^{\star}$ as $\mathbf{W}^{\star}=\mathbf{U{\Sigma}U}^\mathrm{ H}$, with $\mathbf{U}=[\mathbf{u}_1 \cdots \mathbf{u}_M]$.\\
\STATE $p = \Tr(\mathbf{\Sigma})$
\IF {$\Rank ({\mathbf{W}^{\star}})=1$, }
\RETURN  $p^\star= p$ and $\mathbf{w}^{\star} = \mathbf{u}_1$.
\ELSE
\FOR{$d=1,\ldots,D$}
\STATE Generate a random vector $\mathbf{v}_d =\mathbf{U} \mathbf{\Sigma}^{\frac{1}{2}}\mathbf{e}_d$, where $\mathbf{e}_d=[\mathrm{e}^{j{\theta}_1},\mathrm{e}^{j{\theta}_2},...,\mathrm{e}^{j{\theta}_n}]^\mathrm{ H}$ and $\theta_i$ follows the uniform distribution $U(0,2\pi)$.
\IF{ (P3) is feasible with $p$ and $\mathbf{w}_d=\frac{\mathbf{v}_d}{\sqrt{p}}$. }
\RETURN $p^\star= p$ and $\mathbf{w}^{\star}=\mathbf{w}_d$.
\ENDIF
\ENDFOR
\ENDIF
\end{algorithmic}
\end{algorithm}

Different from Algorithm~\ref{Algorithm_rand}, once a feasible solution is obtained for the case $\Rank(\mathbf{W}^\star)\neq1$, Algorithm~\ref{Algorithm_TPM} will end early. The reason is that, in Algorithm~\ref{Algorithm_TPM}, the phase randomization does not affect the value of minimum transmit power $p$ and is used to find a feasible beamforming vector $\mathbf{w}^\star$ that satisfies the constraints of (P3) under a given transmit power.
%

The SDR technique can also be applied to solve the following equivalent problem of (P4).

\textbf{\underline{P4-PSD:}}
\begin{subequations}
\begin{align}
\min_{\mathbf{W}} & \quad\quad \mathrm{Tr}(\mathbf{W}) \\
 \mathrm{s.t.}&\quad\quad \mathbb{E}_c\left[ \log_2\left(1+\frac{\mathbf{H}_\mathrm{eq}(c)\mathbf{W}}{\sigma^2}\right)\right] \geq {\epsilon_s}, \label{eq:P4-PSDC1}\\
  &\quad\quad \frac{\Tr(\mathbf{H}_{2}\mathbf{W})}{\sigma^2} \geq \frac{2^{N \epsilon_c}-1}{N},\\
  &\quad\quad \mathrm{Rank}(\mathbf{W})=1.
\end{align}
\end{subequations}
After solving the SDR of (P4-PSD), we can also obtain a generally approximate beamforming solution $\mathbf{w}^{\star}$ together with the transmit power $p^\star$ to (P4), by using an algorithm analogous to Algorithm~\ref{Algorithm_TPM}. The details are thus omitted for brevity.
\section{low-complexity beamforming optimization} \label{sec:LowC}
Notice that the complexity of solving the formulated problems increases exponentially as the dimension $M$ of the optimization matrix variable $\mathbf{W}$ increases, which may be unaffordable for the case of large-scale antenna array at the PT (i.e., $M\gg1$). In this section, we present a low-complexity beamforming optimization scheme for the considered SR system.

Denote the normalized channel vectors by $\mathbf{\tilde{h}}_1= \frac{\mathbf{h}_1}{{\left\| \mathbf{h}_1 \right\|}}$ and $\mathbf{\tilde{h}}_2= \frac{\mathbf{h}_2}{{\left\| \mathbf{h}_2 \right\|}}$. Then, we have the following proposition.
\begin{proposition}\label{sec:proposition1}
The optimal beamforming vector $\mathbf{w}^{\star}$ for each WSRM or TPM problem has the structure $\mathbf{w}^{\star} = \alpha_1\mathbf{\tilde{h}}_1  + \alpha_2\mathbf{\tilde{h}}_2$, where the complex weights $\alpha_1$ and $\alpha_2$ are subject to $|\alpha_1|^2+|\alpha_2|^2=1$.
\end{proposition}
\begin{IEEEproof}
Please refer to Appendix \ref{proof:pro1}.
\end{IEEEproof}
That is, the optimal beamforming vector $\mathbf{w}^{\star}$ lies in the space spanned by the normalized channel vectors $\mathbf{\tilde{h}}_1$ and $\mathbf{\tilde{h}}_2$.


To demonstrate the advantage of the above beamforming structure, we take (P2) as an example of the WSRM problems and (P3) as an example of the TPM problems.

According to Proposition 3, $\mathbf{w}^{\star}$ can be written as
\begin{equation}\label{eq.w_matrix}
    \mathbf{w}^{\star} = \alpha_1\mathbf{\tilde{h}}_1  + \alpha_2\mathbf{\tilde{h}}_2 = \mathbf{B}\mathbf{a},
\end{equation}
where $\mathbf{B} =[\mathbf{\tilde{h}}_1,\mathbf{\tilde{h}}_2]\in\mathbb C^{M \times 2}$ and $\mathbf{a}=[\alpha_1,\alpha_2]^\mathrm{T} \in\mathbb C^{2\times 1}$.

From~\eqref{eq.w_matrix}, the problem (P2-PSD) can be rewritten as follows

\textbf{\underline{P2-PSD-L:}}
\begin{subequations}
\begin{eqnarray}\label{eq:P2-PSD-L}
\max_{\mathbf{A}}&&\rho\mathbb{E}_c\left[ {{{\log }_2}\left(1 + \frac{\Tr(\mathbf{G}_\mathrm{eq}\mathbf{A})}{\sigma^2}\right)} \right]+  (1-\rho)\frac{1}{N} {{{\log }_2}\left(1 + \frac{N \Tr(\mathbf{G}_{2}\mathbf{A})}{\sigma^2}\right)} \label{P2Lobj}\\
\mathrm{s.t.}&&\Tr(\mathbf{BAB}^\mathrm{H})= p,\label{P2LC1}\\
&&\mathrm{Rank}(\mathbf{A})=1,\label{P2LC2}
\end{eqnarray}
\end{subequations}
where $\mathbf{A}= p \mathbf{aa}^\mathrm{H}\in \mathbb C^{2\times 2}$,
$\mathbf{G}_\mathrm{eq}=\mathbf{B}^\mathrm{H}\mathbf{h}_\mathrm{eq}\mathbf{h}_\mathrm{eq}^\mathrm{H}\mathbf{B}\in\mathbb C^{2\times 2}$ and
$\mathbf{G}_{2}=\alpha |g|^2 \mathbf{B}^\mathrm{H}\mathbf{h}_2\mathbf{h}_2^\mathrm{H}\mathbf{B}\in\mathbb C^{2\times 2}$.
The problem (P2-PSD-L) can also be solved with the SDR technique.

As for the TPM problem, the similar variable transformation is applied. By introducing an additional variable $\mathbf{G}_\mathrm{1}=\mathbf{B}^\mathrm{H}\mathbf{h}_\mathrm{1}\mathbf{h}_\mathrm{1}^\mathrm{H}\mathbf{B}\in\mathbb C^{2\times 2}$, (P3-PSD) is rewritten as follows

\textbf{\underline{P3-PSD-L:}}
\begin{subequations}
\begin{align}
  \min_{\mathbf{A}} & \quad\quad \Tr(\mathbf{BAB}^\mathrm{H})  \\
 \mathrm{s.t.}&\quad\quad \frac{\Tr(\mathbf{G}_{1}\mathbf{A})}{\Tr(\mathbf{G}_{2}\mathbf{A})+\sigma^2} \geq 2^{\epsilon_s}-1, \\
  &\quad\quad \frac{\Tr(\mathbf{G}_{2}\mathbf{A})}{\sigma^2} \geq \gamma_\beta(\epsilon_c),\\
  &\quad\quad \mathrm{Rank}(\mathbf{A})=1.
\end{align}
\end{subequations}
Similar to (P3-PSD), the SDR problem of (P3-PSD-L) can be solved by the CVX.

The variable for both (P2-PSD) and (P3-PSD) is $\mathbf{W}\in \mathbb{C}^{M\times M}$, while the variable for both (P2-PSD-L) and (P3-PSD-L) is $\mathbf{A}\in \mathbb{C}^{2\times 2}$. Compared to (P2-PSD) and (P3-PSD) which optimize the $M$-by-$M$ matrix variable $\mathbf{W}$ directly, (P2-PSD-L) and (P3-PSD-L) only need to optimize a $2$-by-$2$ matrix variable $\mathbf{A}$, thus leading to significantly reduced computational complexity, especially when $M$ is practically large.

%


\section{Simulation Results} \label{sec:NumR}
In this section, simulation results are provided to evaluate the performance of the proposed SR. Independent and identically distributed (i.i.d.) Rayleigh fading is assumed for the direct-link channel $\mathbf{h}_1$ as well as the forward-link channel $\mathbf{h}_2$. The backward-link channel $g$ is assumed to be static, since the BD is typically close to the PR. We define the relative channel gain as $\Delta\Gamma \triangleq \frac{\alpha |g|^2 \mathbb{E}{\left[|{h}_{m,2}|^2\right]}}{\mathbb{E}{\left[|{h}_{m,1}|^2\right]}}$, which mainly depends on the large-scale path loss and the power reflection coefficient $\alpha$. In the simulations, without loss of generality, we set the PT-PR and the PT-BD path loss to be 0 dB, and thus we choose ${h}_{m,i} \sim {\cal{CN}} (0,1)$ for $i=1, 2$. In addition, we define the received SNR as the ratio of transmit power $p$ at the PT and noise power $\sigma^2$ at the PR. The noise power is assumed to be normalized to one. The numerical results are obtained by averaging over $10^4$ channel realizations.

\subsection{Weighted Sum-Rate Maximization}
In this subsection, we consider the WSRM and simulate the rate region performance with different SNR values. Both PSR and CSR setups are considered, and for CSR, we assume $N=128$.



\begin{figure}[t]
    \centering\includegraphics[width=.63\columnwidth]{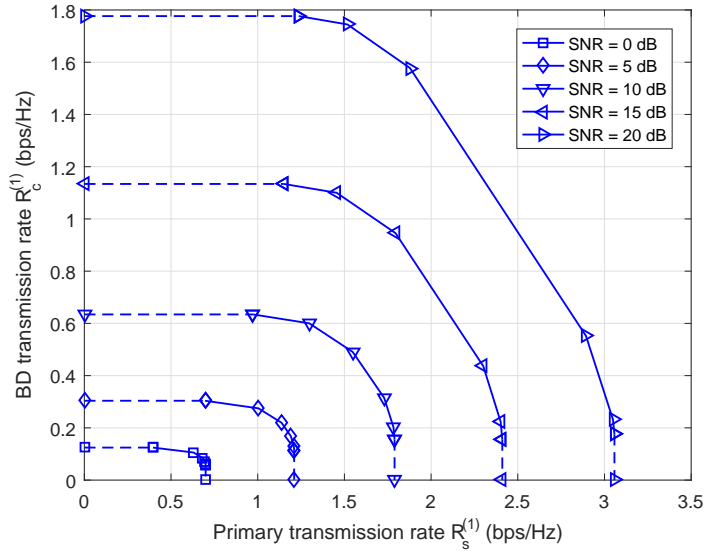}
    \caption{Rate region of the PSR with transmit beamforming: $N=1, \Delta \Gamma = -10 ~\mathrm{dB}$ and $M=2$.}\label{fig:RateRegion}
\end{figure}

\begin{figure}
[t]
  \centering
  \subfigure[Primary and sum transmission rate.]{
    \label{fig:PrimaryRate_20dB} 
    \includegraphics[width=.63\columnwidth]{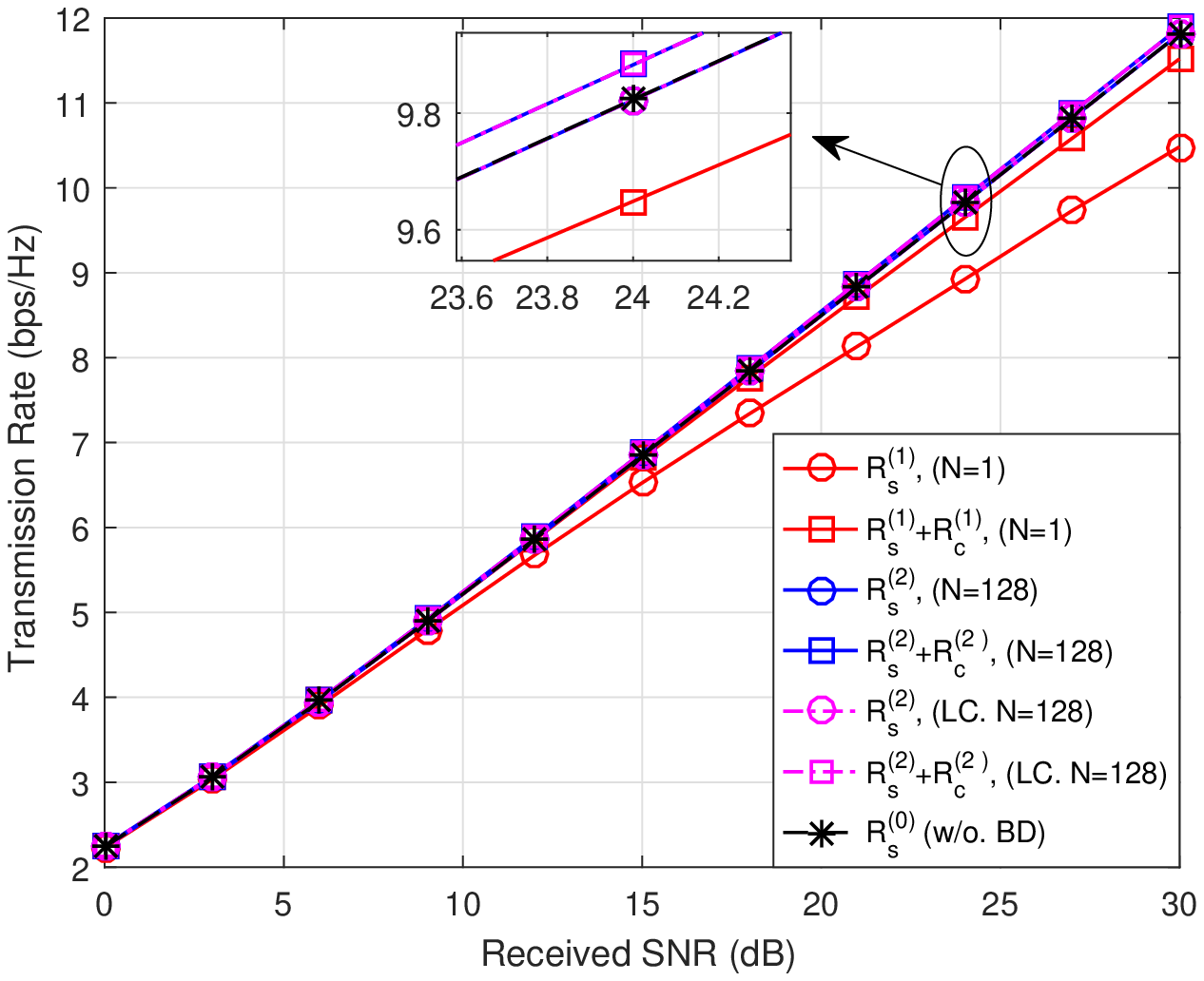}}
  \hspace{1in}
  \subfigure[BD transmission rate.]{
    \label{fig:BDRate_20dB} 
    \includegraphics[width=.63\columnwidth]{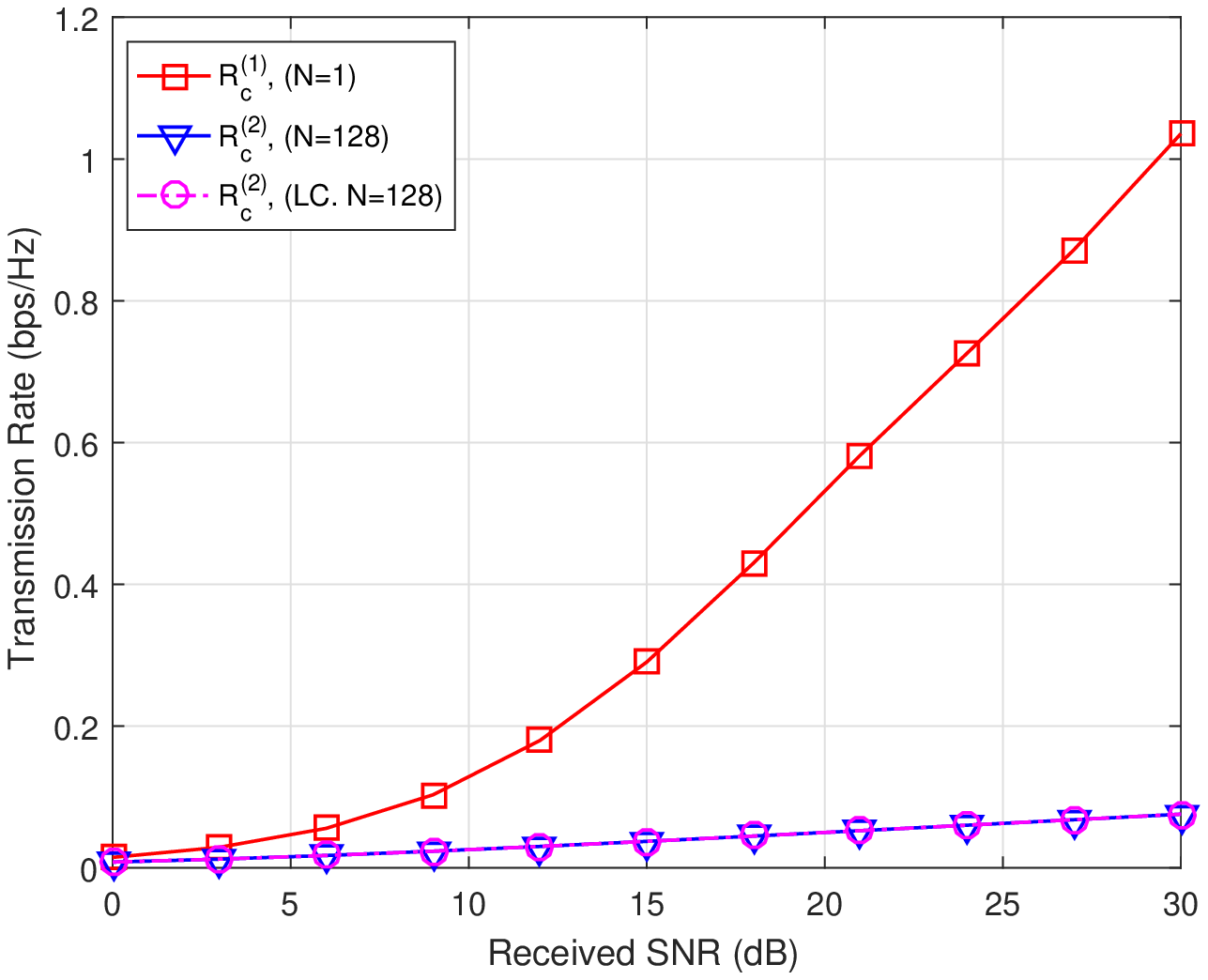}}
  \caption{Rate performances versus the received SNR at PR: $\rho=0.5$, $\Delta \Gamma=-20~\mathrm{dB}$ and $M=4$.}
  \label{fig:SumRate} 
\end{figure}

Fig. \ref{fig:RateRegion} plots the achievable rate regions of the PSR by solving a sequence of WSRM problems with different $\rho$ varying from 0 to 1, for $N=1, \Delta\Gamma=-10 ~\mathrm{dB}$ and $M=2$. Notice that each point of the solid curve represents the pair of maximum primary rate and BD rate by solving (P1) with a given $\rho$, and the dash curve is the extension to the axes. It is observed that the achievable rate region enlarges with the increase of SNR. That is, the increase of SNR can improve both the primary and the BD transmission rates, as expected.




Fig. \ref{fig:SumRate} compares the rate performances versus different received SNRs, when $\rho=0.5, \Delta\Gamma=-20 ~\mathrm{dB}$ and $M=4$. In general, each rate curve increases as the SNR increases. Specifically, for the CSR setup with $N=128$, the system achieves a higher primary rate than that for the PSR setup with $N=1$. This is because that the decoding strategy for CSR exploits the BD signal as a multipath component rather than interference. On the other hand, for CSR, the BD rate is lower than that for the PSR case with $N=1$, due to the longer BD symbol period. In addition, in Fig. \ref{fig:SumRate} the low-complexity method and the conventional method are compared for problem (P2). We observe that by using the low-complexity (LC) beamforming structure, the WSRM problem has almost the same performance as that by using the conventional method.

By comparing Fig. \ref{fig:PrimaryRate_20dB} with Fig. \ref{fig:BDRate_20dB}, it is observed that the primary rate is much higher than the BD rate for each setup, due to the double attenuations in the backscatter link. It is also observed from Fig. \ref{fig:PrimaryRate_20dB} that the CSR system achieves a higher sum rate than the primary system without any BD. Although this sum-rate gain is only moderate, the practical significance of this result lies in that our proposed CSR system enables the backscatter communication concurrently with the primary transmission without any loss in spectral efficiency.
\begin{figure}
[t]
  \centering
  \subfigure[PSR: $N=1$.]{
    \label{fig:Equal_Rc_cst_minP} 
    \includegraphics[width=.63\columnwidth]{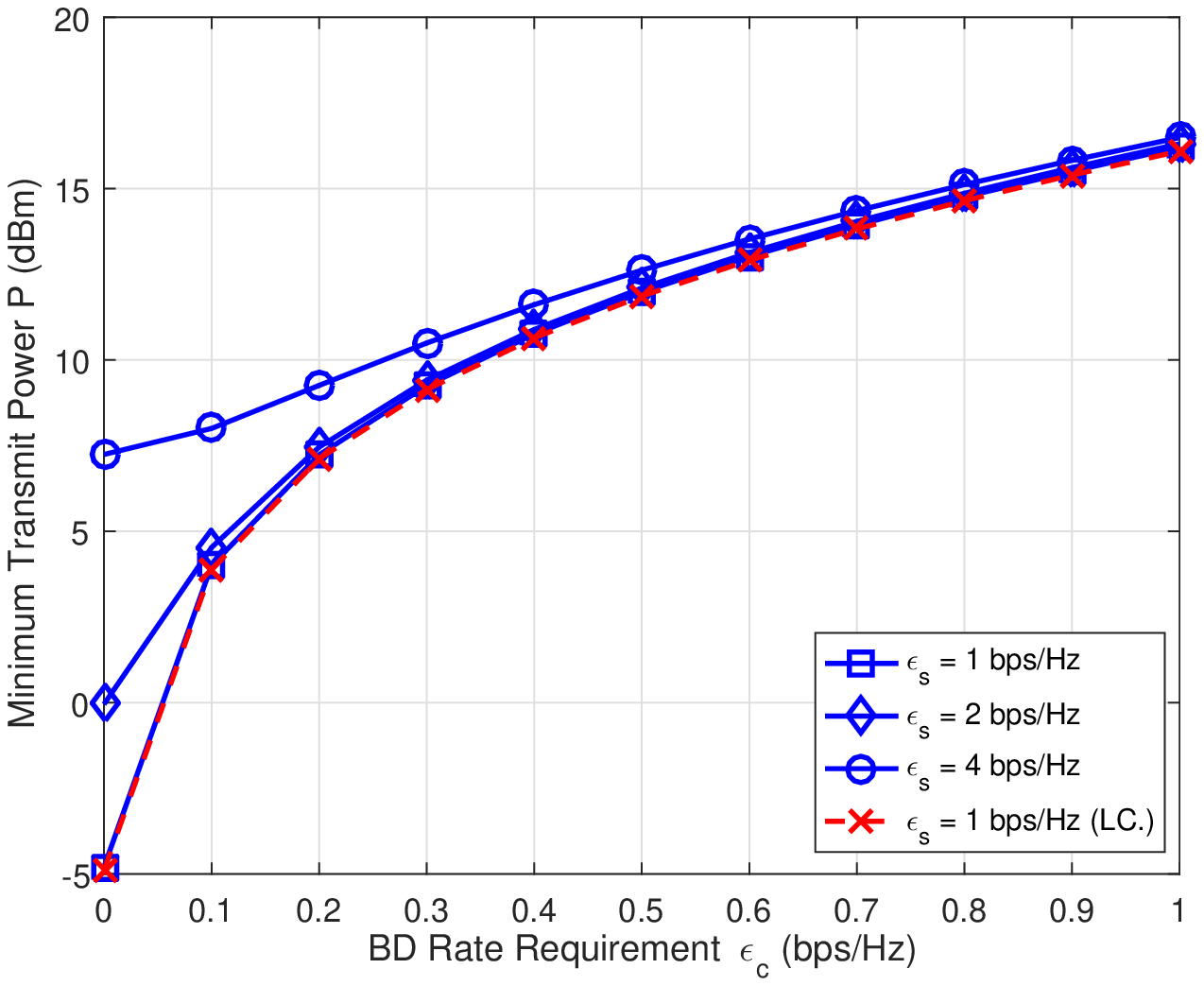}}
  \hspace{1in}
  \subfigure[CSR: $N=128$.]{
    \label{fig:Unequal_Rc_cst_minP} 
    \includegraphics[width=.63\columnwidth]{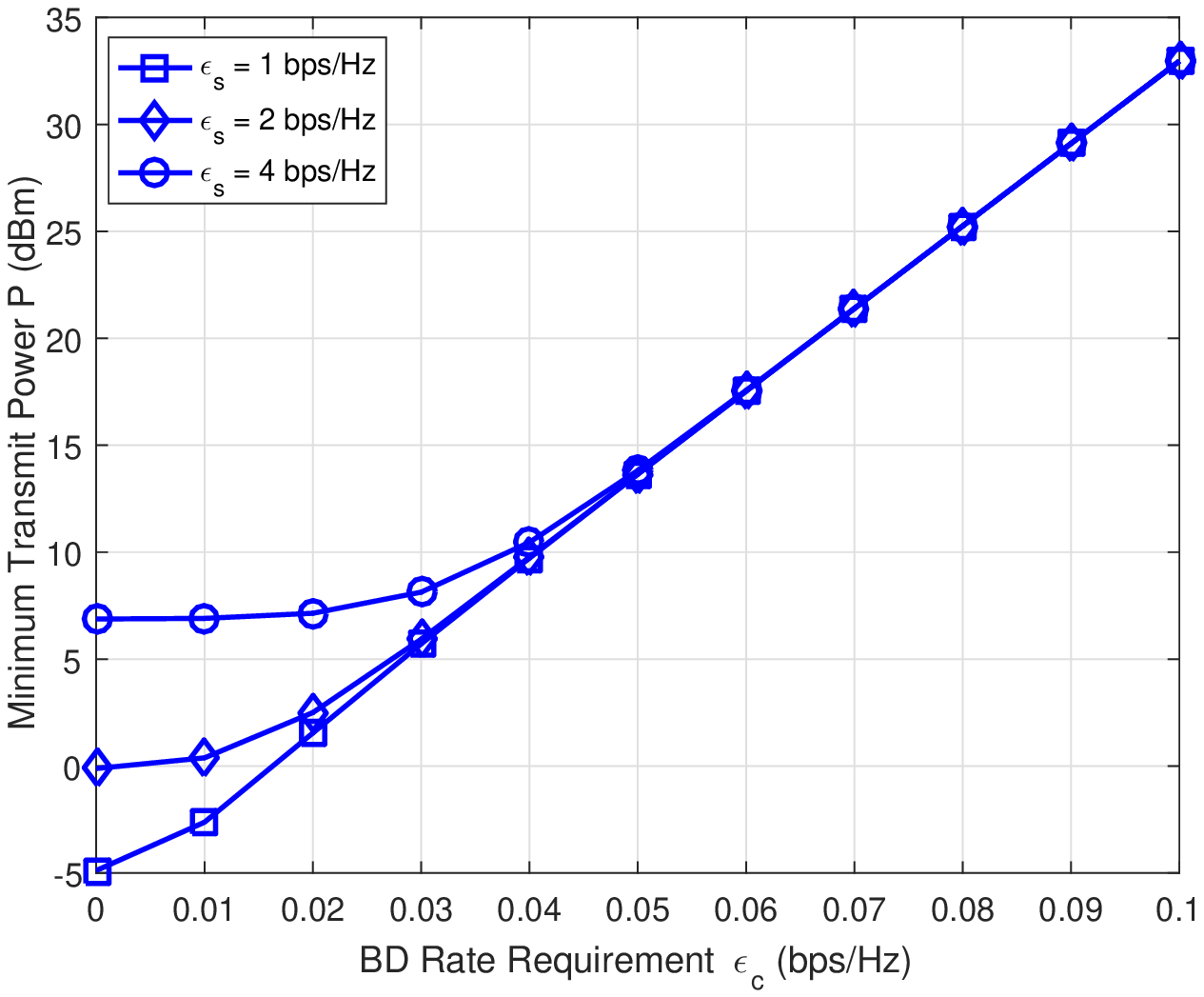}}
  \caption{Minimum transmit power $p$ versus BD rate requirement $\epsilon_c$.}
  \label{fig:MPM_minP} 
\end{figure}


\subsection{Transmit Power Minimization}
In this subsection, we investigate TPM problems under given rate requirements $\epsilon_s$ and $\epsilon_c$ for each setup. For ease of explanation, we generally investigate the TPM problem by varying the BD rate requirement $\epsilon_c$ with a fixed primary rate requirement $\epsilon_s$. We define our transmit power as
\begin{align}\label{eq:powerVsSNR}
  P(\mathrm{dBm}) &= \mathrm{SNR}(\mathrm{dB}) + \mathrm{pathloss}(\mathrm{dB})+\sigma^2(\mathrm{dBm}) \nonumber\\
  &= \mathrm{SNR}(\mathrm{dB}) + \sigma^2(\mathrm{dBm}).
\end{align}

Fig. \ref{fig:Equal_Rc_cst_minP} and Fig. \ref{fig:Unequal_Rc_cst_minP} plot the minimum transmit power versus the BD rate requirement $\epsilon_c$ for PSR ($N=1$) and CSR ($N=128$), respectively. In general, the minimum transmit power increases with the BD rate requirement $\epsilon_c$, but it increases more dramatically for CSR, due to the fact that the rate loss caused by the longer symbol period needs to be compensated with higher transmit power. Also, Fig. \ref{fig:Equal_Rc_cst_minP} shows that the low-complexity method has almost the same performance as the conventional method for TPM problem. Moreover, for both cases, it is observed that for lower $\epsilon_c$, the minimum transmit power increases as the primary rate requirement $\epsilon_s$ increases, but for higher $\epsilon_c$, the minimum transmit power remains the same with different $\epsilon_s$.

\begin{figure}[t]
    \centering\includegraphics[width=.63\columnwidth]{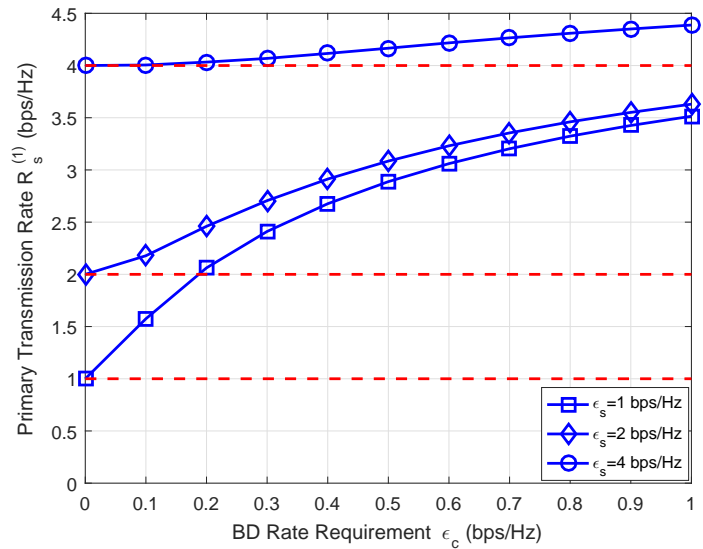}
    \caption{Primary transmission rate $R^{(1)}_s$ versus BD rate requirement $\epsilon_c$: PSR case.}\label{fig:Equal_Rc_cst_Rs}
\end{figure}

\begin{figure}[t]
    \centering\includegraphics[width=.63\columnwidth]{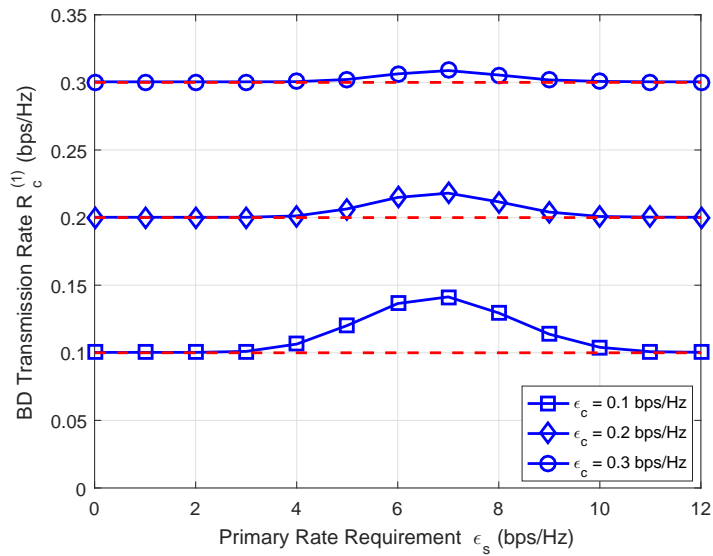}
    \caption{BD transmission rate $R^{(1)}_c$ versus primary rate requirement $\epsilon_s$: PSR case.}\label{fig:Equal_Rs_cst_Rc}
\end{figure}

\begin{figure}[t]
    \centering\includegraphics[width=.63\columnwidth]{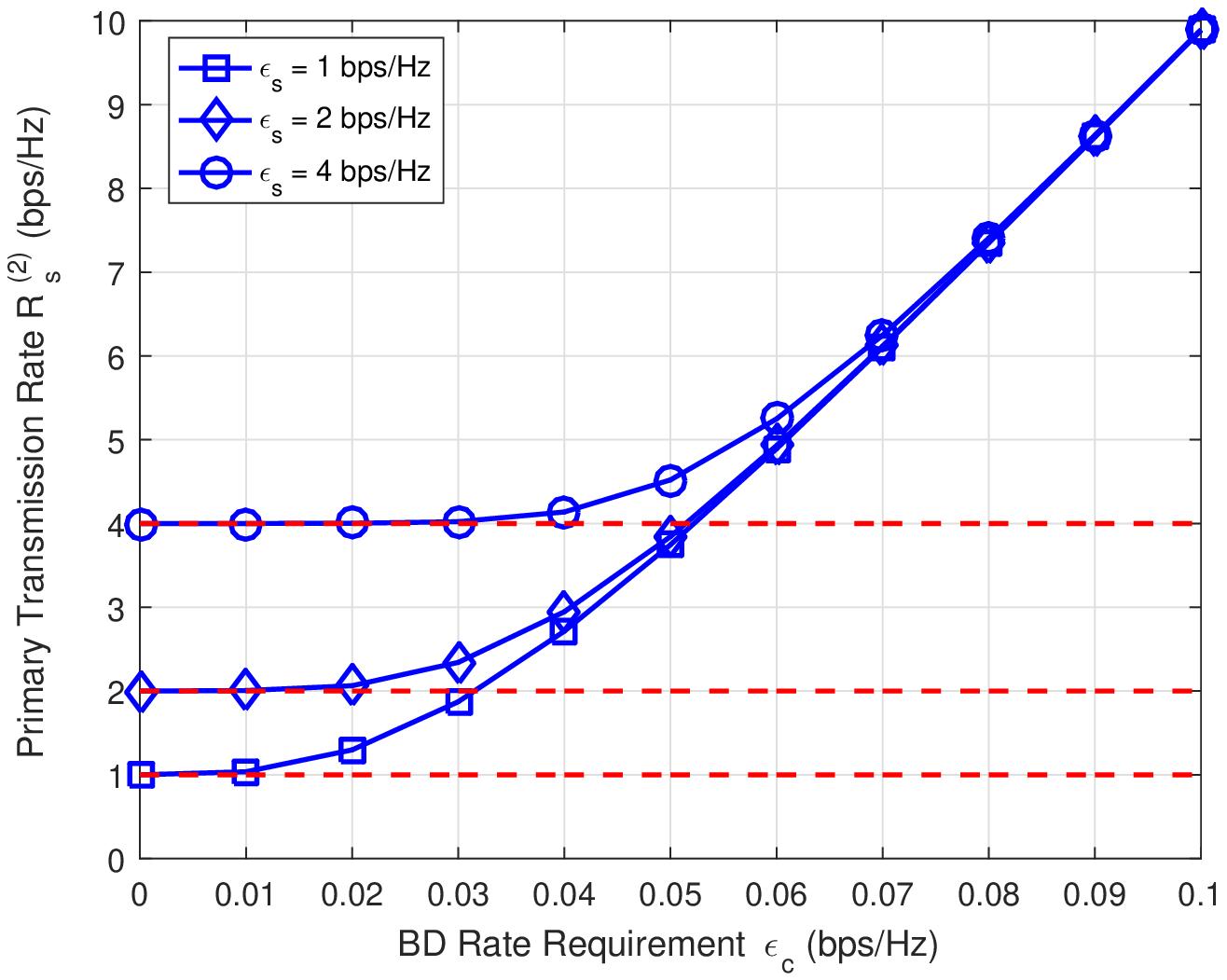}
    \caption{Primary transmission rate $R^{(2)}_s$ versus BD rate requirement $\epsilon_c$: CSR case with $N=128$.}\label{fig:Unequal_Rc_cst_Rs}
\end{figure}
\begin{figure}[t]
    \centering\includegraphics[width=.63\columnwidth]{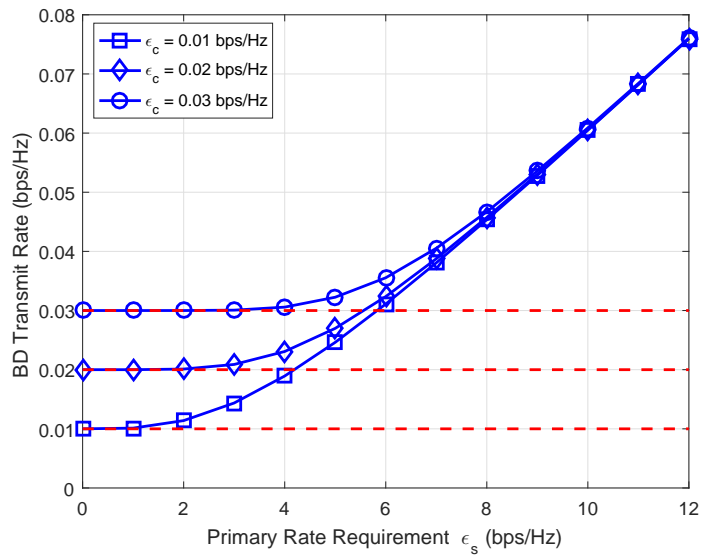}
    \caption{BD transmission rate $R^{(2)}_c$ versus primary rate requirement $\epsilon_s$: CSR case with $N=128$.}\label{fig:Unequal_Rs_cst_Rc}
\end{figure}



Fig. \ref{fig:Equal_Rc_cst_Rs} illustrates the effect of the BD rate requirement $\epsilon_c$ on the primary rate for PSR case. The primary rate $R_s^{(1)}$ increases slowly as the BD rate requirement increases, since higher BD rate requirement results in more transmit power. However, the curve with $\epsilon_s = 4~\mathrm{bps/Hz}$ is flat at first due to the tight primary rate constraint.


Furthermore, we investigate the BD rate performance $R_c^{(1)}$ versus the primary rate requirement $\epsilon_s$. A set of unimodal curves with different $\epsilon_c$ is shown in Fig. \ref{fig:Equal_Rs_cst_Rc}. The phenomenon can be explained as follows. For lower $\epsilon_s$, the BD rate constraint is the bottleneck that limits the transmit power. The BD rate constraint becomes slack as $\epsilon_s$ increases. Since more power is needed to fulfill the primary rate requirement, thus the BD rate increases. However, in high $\epsilon_s$ regime, the transmit power goes to infinity and the primary rate requirement $\epsilon_s$ is only related to the beamforming vector $\mathbf{w}$ as shown below
\begin{equation}\label{eq:MPM_high}
\frac{\left|\mathbf{h}_{1}^{\mathrm{H}}\mathbf{w}\right|^2}{\alpha |g|^2\left|\mathbf{h}_{2}^{\mathrm{H}}\mathbf{w}\right|^2}\geq 2^{\epsilon_s}-1.
\end{equation}
Once the beamforming vector solution is decided by \eqref{eq:MPM_high}, the minimum transmit power depends on the constraint $p\geq{R^{(1)}_{c}}^{-1}(\epsilon_c)/\alpha |g|^2\left|\mathbf{h}_{2}^{\mathrm{H}}\mathbf{w}\right|^2$ and reaches the optimal one when the equality holds. Thus, the BD rate constraint will be tight again.

Similar simulation results are observed for the CSR setup with $N = 128$. Fig. \ref{fig:Unequal_Rc_cst_Rs} shows the effect of the BD rate requirement $\epsilon_c$ on the primary rate $R^{(2)}_s$. As the BD rate requirement $\epsilon_c$ increases, the curves remain unchanged first, then increase gradually and finally coincide with each other. This is due to the fact that, when $\epsilon_c$ is low, the primary rate requirement is the bottleneck that limits the transmit power. As $\epsilon_c$ gradually increases, the BD rate requirement becomes the bottleneck.

In Fig. \ref{fig:Unequal_Rs_cst_Rc}, the BD rate constraint is tight first and then becomes slack as $\epsilon_s$ increases. Compared with Fig. \ref{fig:Equal_Rs_cst_Rc}, the BD rate constraint will not be tight again in Fig. \ref{fig:Unequal_Rs_cst_Rc}. This is due to the fact that the backscattered signal is treated as a multipath component, and there is no interference for the primary transmission in this CSR setup.

\section{Conclusions}\label{sec:Conclusion}

In this paper, a novel technique, called symbiotic radio (SR), has been proposed for passive IoT, in which a backscatter device (BD) is integrated with a primary communication system, and the primary transmitter and receiver are designed to optimize both the primary and BD transmissions. We first present the SIC-based decoding strategy and analyze the achievable rate performance under both PSR and CSR setups. Then, we formulate two problems to maximize the weighted sum rate and minimize transmit power for the considered system, respectively, by optimizing the beamforming vector at the PT. Both problems are recast into equivalent optimization problems with a PSD matrix variable and solved approximately via the technique of SDR. We also propose a novel transmit beamforming structure to reduce the computational complexity of the beamforming optimization. Simulation results show that not only the BD transmission is enabled, but also the primary system achievable rate is improved by exploiting the BD's scattering in the CSR setup.
\appendices
\section{Proof of Proposition 1}\label{proof:pro_Nonchi}
Given channels and transmit beamforming vector, the random variable $T=  \frac{\sqrt{p}}{\sigma}\mathbf{h}_1^\mathrm{ H} \mathbf{w} + \frac{\sqrt{p}}{\sigma}g\mathbf{h}_2^\mathrm{H} \mathbf{w} \sqrt{\alpha}c$, is a linear transformation of a Gaussian random variable $c\sim \mathcal{CN} (0,1)$. Thus we have $T\sim \mathcal{CN}(\frac{\sqrt{p}}{\sigma}\mathbf{h}_1^\mathrm{ H} \mathbf{w},\frac{p \alpha |g|^2 \left|\mathbf{h}^\mathrm{H}_2 \mathbf{w}\right|^2 }{\sigma ^2})$ and its real part and imaginary part are distributed as
\begin{align}
 \mathrm{Re} \left\{T\right\} \sim &\mathcal{N} \left(\mathrm{Re} \left\{\frac{\sqrt{p}}{\sigma}\mathbf{h}_1^\mathrm{ H} \mathbf{w}\right\},\frac{p \alpha |g|^2 \left|\mathbf{h}^\mathrm{H}_2 \mathbf{w}\right|^2 }{2 \sigma ^2} \right), \\
 \mathrm{Im} \left\{T\right\} \sim &\mathcal{N} \left(\mathrm{Im} \left\{\frac{\sqrt{p}}{\sigma}\mathbf{h}_1^\mathrm{ H} \mathbf{w}\right\},\frac{p \alpha |g|^2 \left|\mathbf{h}^\mathrm{H}_2 \mathbf{w}\right|^2 }{2 \sigma ^2} \right).
\end{align}
Since $\mathrm{Re} \left\{T\right\}$ and $\mathrm{Im} \left\{T\right\}$ are independent Gaussian random variables with the same variance $\Sigma = \frac{p \alpha |g|^2 \left|\mathbf{h}^\mathrm{H}_2 \mathbf{w}\right|^2 }{2 \sigma ^2} $, the SNR $\gamma_{s}^{(2)}=\mathrm{Re} \left\{T\right\}^2 + \mathrm{Im} \left\{T\right\}^2$ is distributed as a noncentral chi-square distribution with the non-centrality parameter
\begin{align}
  \lambda &=  \mathrm{Re} \left\{\frac{\sqrt{p}}{\sigma}\mathbf{h}_1^\mathrm{ H} \mathbf{w}\right\}^2+ \mathrm{Im} \left\{\frac{\sqrt{p}}{\sigma}\mathbf{h}_1^\mathrm{ H} \mathbf{w}\right\}^2, \nonumber\\
   &= \frac{p \left|\mathbf{h}^\mathrm{H}_1 \mathbf{w}\right|^2 }{\sigma ^2},
\end{align}
and its PDF is given by
\begin{equation}\label{eq:pdf_rs2Pro}
f(x)= \frac{1}{2\Sigma}\mathrm{e}^{\left(-\frac{x+\lambda}{2\Sigma}\right)}I_0\left(\frac{\sqrt{x\lambda}}{\Sigma}\right).
\end{equation}
The proof is thus completed.

\section{Proof of Proposition 2}\label{proof:pro_Rs}
As the SNR $\gamma_{s}^{(2)}$ is sufficient large, $\log_2(1+\gamma_{s}^{(2)})\simeq \log_2(\gamma_{s}^{(2)})$, thus we have
\begin{align}
{R_{s}^{(2)}} & = \mathbb{E}_c\left[ {{{\log }_2}( \gamma_{s}^{(2)}(c))} \right], \\
   &=\log_2\int_{0}^{\infty}\ln x\frac{1}{2\Sigma}\mathrm{e}^{\left(-\frac{x+\lambda}{2\Sigma}\right)}I_0\left(\frac{\sqrt{x\lambda}}{\Sigma}\right)\mathrm{d}x. \label{eq:Rs_int_m=1}
\end{align}

From \cite{Moser2007Some}, the expected value of the logarithm of a non-central chi-square random variable $V$ with an even number $2m$ of degrees of freedom is given as
\begin{equation}\label{eq:expected_nonchi}
\mathbb{E}\left[\ln V \right] = q_\mathrm{m}(s^2),
\end{equation}
where $s^2$ is the non-centrality parameter, and the function $q_\mathrm{m}(\cdot)$ is defined as follows
\begin{align}\label{eq:qmDefinition}
q_\mathrm{m} \triangleq &\ln(x)- \mathrm{Ei}(-x)+\sum_{j=1}^{m-1}(-1)^j\left[e^{-x}(j-1)!-\frac{(m-1)!}{j(m-1-j)!}\right]\left(\frac{1}{x}\right)^j, x>0
\end{align}

%

That is
\begin{equation}\label{eq:intSolution}
\int_{0}^{\infty}\ln v\cdot\left(\frac{v}{s^2}\right)^{\frac{m-1}{2}}e^{-v-s^2}I_{m-1}\left(2s\sqrt{v}\right)\mathrm{d}v=q_m(s^2),
\end{equation}
for any $m\in\mathbb{N}$ and $s^2\geq0$. Applying the linear transformation $v=\frac{x}{2\Sigma},s^2 = \frac{\lambda}{2\Sigma}$ to \eqref{eq:Rs_int_m=1}, we have
\begin{subequations}
\begin{align}
   {R_{s}^{(2)}}& = \log_2\mathrm{e} \int_{0}^{\infty} \ln v \mathrm{e}^{(-v-s^2)}I_0\left(2s\sqrt{v}\right) \mathrm{d}v \nonumber \\
   &~~~+\log_2\left(2\Sigma\right)\int_{0}^{\infty}\mathrm{e}^{(-v-s^2)}I_0\left(2s\sqrt{v}\right) \mathrm{d}v, \\
   &=\log_2\mathrm{e} \cdot q_1(s^2)+\log_2(2\Sigma),\label{eq:pdf_int}\\
   &=\log_2\mathrm{e}\cdot q_1\left(\frac{\lambda}{2\Sigma}\right) + \log_2(2\Sigma), \\
   &= \log_2 \lambda- \mathrm{Ei}\left(-\frac{\lambda}{2\Sigma}\right) \log_2 \mathrm{e}.
\end{align}
\end{subequations}
Equation \eqref{eq:pdf_int} is due to the fact that the second term is an integral over a noncentral chi-square distribution. Thus, the proof is completed.

\section{Proof of Proposition 3}\label{proof:pro1}
Let the beamforming vector be
\begin{equation}\label{eq:w}
  \mathbf{w}= \sum\limits_{{i} = 1}^{ 2} {{\alpha _i}\tilde{\mathbf{h}}_i}+\sum\limits_{{j} = 1}^{{M - 2}} {{\eta _j}\tilde {\mathbf{t}}_j^{\perp}},
\end{equation}
where $\tilde {\mathbf{t}}_j^{\perp}$ is the basis of the null space of $\{ \tilde{\mathbf{h}}_i \}$, i.e., $\tilde{\mathbf{h}}_i^\mathrm{H}\tilde {\mathbf{t}}_j^{\perp}=0$. It can be verified that $\tilde {\mathbf{t}}_j^{\perp}$ cannot contribute to improve the SNR in the objective functions of WSRM problems (P1) and (P2), while $\tilde{\mathbf{h}}_i,i=1,2$ can help to improve the SNR in the objective function.

For the SINR expression in the objective function in (P1), the beamforming vector satisfies the condition that $\mathbf{w}=a\tilde{\mathbf{h}}_1 + b\tilde{\mathbf{h}}^\perp_2+\sum\limits_{{j} = 1}^{{M - 2}} {{\eta _j}\tilde {\mathbf{t}}_j^{\perp}}$, where $\tilde{\mathbf{h}}_2^\mathrm{H}\tilde {\mathbf{h}}_2^{\perp}=0$,
$\tilde{\mathbf{h}}_1^\mathrm{H}\tilde {\mathbf{h}}_2^{\perp} > 0 $,
and $\tilde {\mathbf{h}}_2^{\perp}=x\tilde{\mathbf{h}}_1+y\tilde{\mathbf{h}}_2$, thus $\mathbf{w}$ satisfies the structure~\eqref{eq:w}.
It is easy to verify that $\tilde{\mathbf{h}}_1$ and $\tilde{\mathbf{h}}_2^{\perp}$ help improve the SINR of the primary transmission while $\tilde {\mathbf{t}}_j^{\perp}$ does not.
Since the component $\tilde {\mathbf{t}}_j^{\perp}$ cannot contribute to improve the value of the objective function, we only need to optimize the coefficients of $\tilde{\mathbf{h}}_1$ and $\tilde{\mathbf{h}}_2$ to find the optimal beamforming vector $\mathbf{w}^{\star}$.

In addition, since the beamforming vector $\mathbf{w}^{\star}$ is a normalized one, the complex weights $\alpha_1$ and $\alpha_2$ are subject to $|\alpha_1|^2+|\alpha_2|^2=1$.

Since the TPM problems (P3) and (P4) have the same SNR or SINR expressions, the same results hold for the TPM problems. The proof is thus completed.

\renewcommand{\baselinestretch}{1}
\bibliographystyle{IEEEtran}
\bibliography{IEEEabrv,library_bac}
\end{document}